\documentclass{PoS}

\usepackage{amsmath}
\usepackage[boxruled,vlined,linesnumbered]{algorithm2e}

\SetNlSkip{.9em}
\SetCommentSty{emph}
\SetArgSty{text}
\SetAlFnt{\small}

\DeclareMathOperator\ceil{ceil}
\DeclareMathOperator\diag{diag}
\newcommand\WMG{\mbox{DD-$\alpha$AMG}}
\newcommand\Ntv{N_\text{tv}}
\newcommand\NS{N_\text{SIMD}}

\title{Adaptive algebraic multigrid on SIMD architectures\thanks{Work
    supported by the German Research Foundation (DFG) in the framework
    of SFB/TRR-55.}} 
\ShortTitle{DD-$\alpha$AMG on SIMD architectures}

\author{Simon Heybrock$^{a,b}$, Matthias Rottmann$^c$, Peter Georg$^a$, 
  \speaker{Tilo Wettig}$\,^a$\\
  $^a$Department of Physics, University of Regensburg, 93040
  Regensburg, Germany\\
  $^b$Data Management and Software Centre, European Spallation Source, 
  Universitetsparken 5, 2100 Copenhagen, Denmark\\
  $^c$Department of Mathematics, University of Wuppertal, 42097
  Wuppertal, Germany\\ 
  E-mail: \email{tilo.wettig@ur.de}}

\abstract{We present details of our implementation of the Wuppertal
  adaptive algebraic multigrid code \WMG\ on SIMD architectures, with
  particular emphasis on the Intel Xeon Phi processor (KNC) used in
  QPACE~2.  As a smoother, the algorithm uses a
  domain-decomposition-based solver code previously developed for the
  KNC in Regensburg.  We optimized the remaining parts of the
  multigrid code and conclude that it is a very good target for SIMD
  architectures.  Some of the remaining bottlenecks can be eliminated
  by vectorizing over multiple test vectors in the setup, which is
  discussed in the contribution of Daniel Richtmann.}

\FullConference{The 33nd International Symposium on Lattice Field Theory}

\begin{document}

\section{Introduction}
\vspace*{-2mm}

In order to solve the Dirac equation $Du=f$ on large lattices we
usually use an iterative method and a preconditioner $M$ with
$M^{-1}\approx D^{-1}$.  Writing $DM^{-1}Mu=DM^{-1}v=f$ we first solve
for $v$ with the preconditioned matrix $DM^{-1}$, which is much better
behaved than $D$, and then obtain $u=M^{-1}v$.  In each iteration the
residual $r_n=f-DM^{-1}v_n$ gives an indication of how close we are to
the true solution.  Formally, the residual can be written as a linear
combination of the eigenmodes of $D$.  The art is to find a
preconditioner that eliminates, or at least reduces, the contributions
of these eigenmodes to the residual.  Adaptive algebraic multigrid
(MG) \cite{Babich:2010qb} is such a preconditioner, and here we focus
on the Wuppertal version \WMG\ \cite{Frommer:2013fsa}.  It consists of
two parts, a coarse-grid correction (CGC) that reduces the
contributions of the low modes and a domain-decomposition (DD) based
smoother that reduces the contributions of the high modes
\cite{Luscher:2003qa}.

In this contribution we implement and optimize the Wuppertal code for
SIMD architectures.\footnote{We restrict ourselves to an MG V-cycle
  with two levels (coarse grid and fine grid), for reasons explained
  in Sec.~\ref{sec:implementation}.}  Our main target is the Intel
Xeon Phi processor (KNC) used in QPACE~2, but our code can easily be
adapted to other SIMD-based architectures.  For the smoother we use
the code already developed for the KNC in Regensburg
\cite{Heybrock:2014iga}.  Here we mainly focus on the remaining parts
of the code.

\vspace*{-2mm}
\section{Description of the algorithm}
\vspace*{-2mm}

Some preliminaries: The lattice volume $V$ is divided into
$N_\text{block}$ blocks.  To implement \mbox{$\gamma_5$-sym}\-metry
\WMG\ defines, for each block, two aggregates that contain the left-
and right-handed spinor components of the fields,
respectively.\footnote{I.e., an aggregate contains $6V_\text{block}$
  degrees of freedom, which we can enumerate with an index
  $n=1,\ldots,6V_\text{block}$.}  The outer Dirac solver is FGMRES
\cite{Saad2003iterative}.  The MG method consists of two parts: (1)
the initial setup phase and (2) the application of the MG
preconditioner (Alg.~\ref{alg:precond}) in every FGMRES iteration.
For pedagogical reasons we first explain the latter.

\vspace*{-2mm}
\subsection{MG preconditioner}
\vspace*{-2mm}

\begin{algorithm}[b]
  \caption{MG preconditioner (V-cycle)}
  \label{alg:precond}
  \DontPrintSemicolon 
  \KwIn{right-hand side $y$}
  \KwOut{approximate solution $x$ of $Dx=y$}
  apply coarse-grid correction to $y$ (Alg.~\ref{alg:CGC})\;
  apply smoother to $y$, with result from coarse-grid correction as starting guess (Alg.~\ref{alg:DD})\;
  set $x$ to result of smoother
\end{algorithm}

\begin{algorithm}
  \caption{Coarse-grid correction}
  \label{alg:CGC}
  \DontPrintSemicolon 
  \KwIn{right-hand side $y$}
  \KwOut{approximate solution $x$ of $Dx=y$}
  \tcp*[h]{Restrict (i.e., project):}\;
  restrict vector $y$ from fine to coarse grid: \hfill
  \tcp*[h]{$\dim(R)=2N_\text{\rm tv}N_\text{\rm block}\times12V$}
  \begin{align}
    \label{eq:R}
    y_c=Ry\quad\text{with }
    R=\diag(R_1^\ell,R_1^r,\ldots,
    R_{N_\text{block}}^\ell,R_{N_\text{block}}^r)
    \text{ and } R_i^{\ell/r} \text{ from Alg.~\ref{alg:pcgo}}
  \end{align}
  \tcp*[h]{Coarse-grid solve to low precision using FGMRES with
    even/odd preconditioning: $x_c\approx D_c^{-1}y_c$}\; 
  \Repeat{norm of residual $\lesssim0.05$}{
    apply coarse-grid operator to current iteration vector\;
    BLAS-like linear algebra (mainly Gram-Schmidt)\;
  }
  \tcp*[h]{Prolongate (i.e., interpolate):}\;
  extend solution vector from coarse to fine grid: $x=P x_c$ with
  $P=R^\dagger$ 
\end{algorithm}

In the coarse-grid correction (Alg.~\ref{alg:CGC}), we first restrict
the current iteration vector of the outer solver from a fine to a
coarse grid using a restriction operator that approximately preserves
the low modes of the Dirac operator.  How this operator is constructed
is explained in Sec.~\ref{sec:setup}.  The Dirac equation is then
solved to low precision on the coarse grid, and the solution is lifted
back to the fine grid.  The main point is that this solution
approximates the low-mode content of the true solution.

\begin{algorithm}
  \caption{Smoother (DD)}
  \label{alg:DD}
  \DontPrintSemicolon 
  \KwIn{right-hand side $y$, starting guess $x^0$}
  \KwOut{approximate solution $x^{(N_\text{\rm smoother})}$
    of $Dx=y$}
  split lattice into blocks\;
  write $D=B+Z$ with $B=$ couplings within blocks and $Z=$ couplings between blocks\;
  \For{$n=1$ \KwTo $N_\text{\rm smoother}$}{
    $x^{(n)}=x^{(n-1)}+B^{-1}(y-Dx^{(n-1)})$ 
    \hfill\tcp*[h]{simplified; in practice SAP \cite{Luscher:2003qa} is used}
  }  
\end{algorithm}

To also approximate the high-mode content of the true solution, a
DD-based smoother (Alg.~\ref{alg:DD}) is applied to the current
iteration vector.  The inverse block size acts as a cutoff for the low
modes.  Thus, for vectors $y_\text{high}$ that do not contain modes
below this cutoff, the smoother applied to $y_\text{high}$
approximates $D^{-1}y_\text{high}$.  If we use the solution from the
coarse-grid correction (which approximates the low modes of the true
solution) as a starting guess $x^0$, the smoother works on
$y_\text{high}=y-Dx^0$, i.e., the low modes have been approximately
eliminated from $y$ by the subtraction of $Dx^0$.

\vspace*{-2mm}
\subsection{MG setup}
\label{sec:setup}
\vspace*{-2mm}

In any MG method one needs to restrict from a fine to a coarse grid.
In geometric MG, the restriction proceeds by simply averaging subsets
of the fields on the fine grid.  Algebraic MG is more sophisticated
and includes nontrivial weight factors in the average.  The art is to
find good weight factors so that the low modes of the operator to be
inverted are approximately preserved after the
restriction.\footnote{This is the same principle as inexact deflation
  \cite{Luscher:2007se}.}  The purpose of the MG setup phase is the
computation of these weight factors.  The main idea is to apply an
iterative process through which more and more of the high-mode
components are eliminated.  To this end, we define a set of test
vectors $\{ v_j \, : \, j=1,\ldots,N_{tv} \}$ (each of dimension
$12V$) that will, at the end of the iterative process, approximate the
low-mode components.  The iterative process is described in
Alg.~\ref{alg:setup}, see \cite{Frommer:2013fsa} for details.
Alg.~\ref{alg:pcgo} describes how the restriction operator is
constructed from the test vectors $v_j$ and how $D$ is restricted to
the coarse grid.

\begin{algorithm}[t]
  \caption{MG setup}
  \label{alg:setup}
  \DontPrintSemicolon
  \tcp*[h]{Initial setup:}\;
  set the $\Ntv$ test vectors to random starting vectors\;
  \For{$k=1$ \KwTo $3$}{
    update each test vector by applying smoother with $N_\text{\rm
      smoother}=k$, with
    starting guess 0 (Alg.~\ref{alg:DD})\;
  }
  setup of restriction and coarse-grid operator (Alg.~\ref{alg:pcgo})\;
  normalize the test vectors\;
  \tcp*[h]{Iterative setup:}\;
  \For{$i=1$ \KwTo $N_\text{setup}$}{
    \For{$j=1$ \KwTo $\Ntv$}{
      apply CGC to test vector $v_j$ (Alg.~\ref{alg:CGC})\;
      apply smoother to test vector $v_j$, with
      result from CGC as starting guess (Alg.~\ref{alg:DD})\;
      replace test vector $v_j$ by result of smoother\;
    }
    setup of restriction and coarse-grid operator (Alg.~\ref{alg:pcgo})\;
  }
\end{algorithm}

\begin{algorithm}
  \caption{Setup of restriction and coarse-grid operator}
  \label{alg:pcgo}
  \DontPrintSemicolon 
  \KwIn{test vectors $\{v_j\}$}
  \KwOut{restriction operator $R$ and coarse-grid operator $D_c$}
  \tcp*[h]{Setup of restriction operator:}\;
  \For{$i=1$ \KwTo $N_\text{\rm block}$}{
    \ForEach{$h=\ell,r$}{
      set $R_i^h$ to $\Ntv\times6V_\text{\rm block}$
      matrix having in its rows the vectors $v_j^\dagger$ restricted to aggregate $A_i^h$\;
      run Gram-Schmidt on the rows of $R_i^h$\;
    }
  }
  \tcp*[h]{Setup of coarse-grid operator by restriction:}\;
  \For{$i=1$ \KwTo $N_\text{\rm block}$}{
    \ForEach(\tcp*[f]{optimize using \eqref{eq:dagger}})
    {$j\in\{i\text{ and nearest neighbors of }i\}$}{ 
      compute couplings between sites $i$ and $j$ on coarse grid:
      \hfill\tcp*[h]{no sums over $i,j,\ell,r$}
      \begin{align}
        \label{eq:cgo}
        \vspace*{-2mm}
        \begin{pmatrix}
          D_c^{\ell\ell} & D_c^{\ell r}\\
          D_c^{r\ell} & D_c^{rr}
        \end{pmatrix}_{ij}
        =
        \begin{pmatrix}
          R_i^\ell & 0\\
          0 & R_i^r
        \end{pmatrix}
        \begin{pmatrix}
          D_{ij}^{\ell\ell} & D_{ij}^{\ell r}\\
          D_{ij}^{r\ell} & D_{ij}^{rr}
        \end{pmatrix}
        \begin{pmatrix}
          P_j^\ell & 0\\
          0 & P_j^r
        \end{pmatrix}\\[-7mm]\notag
      \end{align}
    }
  }
  \tcp*[h]{$\dim(D_c^{hh'})_{ij}=N_\text{\rm tv}$,
    $\dim(D_{ij}^{hh'})=6V_\text{\rm block}$}\;
  \tcp*[h]{\eqref{eq:cgo} can be written as
    $(D_c)_{ij}^{hh'}=R_i^hD_{ij}^{hh'}P_j^{h'}$ or as
    $D_c=RDP$ with $R$ defined in \eqref{eq:R} and $P=R^\dagger$.}
\end{algorithm}

\vspace*{-2mm}
\section{Implementation details}
\label{sec:implementation}
\vspace*{-2mm}

The implementation of the DD-based smoother (Alg.~\ref{alg:DD}) on a
SIMD architecture is quite complicated and described in detail in
\cite{Heybrock:2014iga}.  In contrast, the remaining parts of the
\WMG\ code are easier to vectorize since the number of components that
can be treated on the same footing contains a factor of $\Ntv$ (on the
fine grid) or $2\Ntv$ (on the coarse grid).  If we choose this factor
to be equal to an integer multiple of the SIMD length $\NS$ (i.e., the
number of SIMD components) we achieve perfect use of the SIMD
unit.\footnote{For the KNC, $\NS=16$ in single precision.  We use
  $\Ntv=16$ to $32$.  This choice is appropriate from an algorithmic
  point of view.  Should the algorithmic performance require $\Ntv$ to
  be unequal to an integer multiple of $\NS$ we would use
  $\ceil(\Ntv/\NS)$ SIMD vectors on the fine grid and pad the last of
  these (and the corresponding memory region) with zeros.  Analogously
  for the coarse grid.  E.g., for $\Ntv=24$ we need padding only on
  the fine grid.\label{footnote}} On the downside, \WMG\ consists of
many parts, most of which take a non-negligible part of the total
execution time.  Therefore we need to vectorize/optimize all of them.
Before going into detail we briefly summarize general aspects of our
work.

We streamlined the original Wuppertal code and removed some redundant
or unnecessary parts (e.g., some orthonormalizations).  We reduced the
memory consumption slightly by eliminating redundancies and temporary
copies.  We threaded the code by decomposing the lattice into pieces
that are assigned to individual threads.  Based on earlier
microbenchmarks, we decided to use persistent threads with
synchronization points, rather than a fork-join model.  Our SIMD
implementation is based on intrinsics for the Intel compiler.  To
facilitate efficient SIMD multiplication of complex numbers, the data
layout of $R$ and $D_c$ is such that real and imaginary parts are not
mixed in the same register.  In this paper we only describe 2-level
MG, where all KNCs work both on the fine and the coarse grid.  Many
parts carry over to 3-level (or higher), with two exceptions: (1)
since the KNC has rather limited memory, more levels lead to more idle
cores or even idle processors, and (2) we currently do not have an
efficient implementation of the DD smoother on a coarse
grid.\footnote{For two levels this operator is not needed.  For higher
  levels we cannot reuse our fine-grid DD smoother since it is for a
  different operator.}

We now describe how the key components that needed to be optimized
were implemented.

\textbf{Restriction (Alg.~\ref{alg:CGC}):} We vectorize the
matrix-vector multiplication \eqref{eq:R} for each aggregate as
described in Alg.~\ref{alg:Ry}.  The vectorization is done such that
the row index of $R_i^{\ell/r}$ runs in the SIMD vector, i.e., the
latter contains a column of $R_i^{\ell/r}$ if $\Ntv=\NS$.\footnote{For
  $\Ntv\ne\NS$ footnote~\ref{footnote} applies.  The need for padding
  on the fine grid could be reduced by combining $R_i^\ell$ and
  $R_i^r$ at the expense of a more complicated broadcast in
  Alg.~\ref{alg:Ry}.  This is a tradeoff between memory bandwidth and
  instruction count that we have not explored yet.}

\begin{algorithm}[t]
  \caption{SIMD implementation of $Ry$ in \protect\eqref{eq:R}}
  \label{alg:Ry}
  \DontPrintSemicolon 
  \For{$i=1$ \KwTo $N_\text{block}$}{
    \ForEach{$h=\ell,r$}{
      set $(y_c)_i^h=0$ in SIMD vectors (real and imaginary part) 
      \hfill\tcp*[h]{$\dim(y_c)_i^h=N_\text{\rm tv}$\hspace*{-6.6mm}}\;
      \For(\tcp*[f]{work on aggregate $A_i^h$})
      {$n=1$ \KwTo $6V_\text{block}$}{
        load real and imaginary part of column $n$ of $R_i^h$ into SIMD vectors\;
        broadcast real and imaginary part of corresponding element of $y$
        into SIMD vectors\;
        increase $(y_c)_i^h$ by complex fused multiply-add
        (corresponding to 4 real SIMD fmadds)\;
      }
      write $(y_c)_i^h$ to memory\;
    }
  }
\end{algorithm}

\textbf{Prolongation (Alg.~\ref{alg:CGC}):} Similar to restriction but
with $R\to P=R^\dagger$.  Since the aspect ratio of the rectangular
matrix is reversed, now the column index of $P_i^{\ell/r}$ ($=$ row
index of $R$) runs in the SIMD vector.  At the end an additional sum
over the elements in the SIMD vector is required, which makes
prolongation somewhat less efficient than restriction.

\begin{algorithm}
  \caption{SIMD implementation of $D_{ij}^{hh'}P_j^{h'}$
    in \protect\eqref{eq:cgo}.}
  \label{alg:DP}
  \DontPrintSemicolon 
  \For{$x\in\text{ block }i$}{
    set output $=0$ in SIMD vectors (real and imaginary parts)\;
    \ForEach{$\mu\in\{\pm1,\pm2,\pm3,\pm4\}$}{
      \If{$x+\hat\mu\in\text{ block }j$}{
        \mbox{load real and imag. parts of the 6 rows of $P_j^{h'}$
        corresponding to $x\!+\!\hat\mu$ into SIMD vectors\hspace*{-10mm}}\;
        broadcast real and imag. parts of the 9 elements of SU(3) link $U_\mu(x)$
        into SIMD vectors\;
        increase output by complex fmadd
        $(1+\gamma_\mu)^{hh'}U_\mu(x)^\dagger P_j^{h'}(x+\hat\mu)$ \;
      }
    }
    \If{$i=j$ and $h=h'$}{
      load real and imaginary parts of the 6 rows of $P_i^{h}$
      corresponding to $x$ into SIMD vectors\;
      broadcast real and imaginary parts of the clover matrix elements
      $C^{hh}(x)$ into SIMD vectors\; 
      increase output by complex fmadd $C^{hh}(x)P_i^{h}(x)$ \;
    }
  }
\end{algorithm}

\textbf{Setup of coarse-grid operator (Alg.~\ref{alg:pcgo}):} In
\eqref{eq:cgo} we first compute $D_{ij}^{hh'}P_j^{h'}$, which
corresponds to the application of the sparse matrix $D_{ij}^{hh'}$ to
multiple vectors, i.e., the columns of $P_j^{h'}$.  The vectorization
of this operation is described in Alg.~\ref{alg:DP}.  The application
of $R$ to the result corresponds to a restriction with multiple
right-hand sides, which can be optimized as described in the
contribution of Daniel Richtmann \cite{Richtmann:lat15}.  As in the
original Wuppertal code, we only compute \eqref{eq:cgo} for $i=j$ and
the forward neighbors.  For the backward neighbors we use
\vspace*{-2mm}
\begin{align}
  \label{eq:dagger}
  (D_c)^{hh}_{ji}=(D_c)^{hh\dagger}_{ij}\quad\text{and}\quad
  (D_c)^{hh'}_{ji} = - (D_c)^{h'h\dagger}_{ij}\;(h\ne h')\,.
  \notag\\[-8mm]
\end{align}
In contrast to the original code we also store $(D_c)_{ji}$ since a
transpose is expensive in SIMD (i.e., we perform the transpose only
when the operator is constructed, but not when it is applied).  The
coarse-grid operator can be stored in half precision to reduce working
set and memory bandwidth requirements.  We did not observe any
negative effects of this on stability or iteration count and therefore
made it the default.

\textbf{Application of coarse-grid operator (Alg.~\ref{alg:CGC}):} The
spatial structure of $D_c$ is similar to Wilson clover.  In
\eqref{eq:cgo}, $(D_c)_{ij}\ne0$ only if $i$ and $j$ are equal or
nearest neighbors.  In that case $(D_c)_{ij}$ is dense and stored in
memory.  Therefore the vectorization can be done as in the
restriction.

\textbf{Gram-Schmidt on aggregrates (Alg.~\ref{alg:pcgo}):} We do not
use modified Gram-Schmidt \cite{Saad2003iterative} since (1) classical
Gram-Schmidt lends itself more easily to vectorization and needs fewer
globals sums and (2) stability of the Gram-Schmidt process is not
expected to become an issue in the preconditioner.  To obtain better
cache reuse and thus save memory bandwidth we use the block
Gram-Schmidt method \cite{Jalby:1991}.  As usual, the vectorization is
done by merging the same components of the $\Ntv$ test vectors in the
SIMD vectors.  The disadvantage of this strategy is that axpy
operations and dot products then waste parts (on average one half) of
the SIMD vectors.

\textbf{BLAS-like linear algebra on coarse grid (Alg.~\ref{alg:CGC}):}
In order to utilize the SIMD unit we would need to change the data
layout on the coarse grid.  This change would propagate to other parts
of the code.  We have not yet made this change since it requires
nontrivial programming efforts, while the impact on performance is not
dominant.  As a temporary workaround we sometimes de-interleave real
and imaginary parts on the fly to do a SIMD computation.

\vspace*{-2mm}
\section{Performance and conclusions}
\vspace*{-2mm}

\begin{figure}
  \centering
  \includegraphics[width=.95\textwidth]{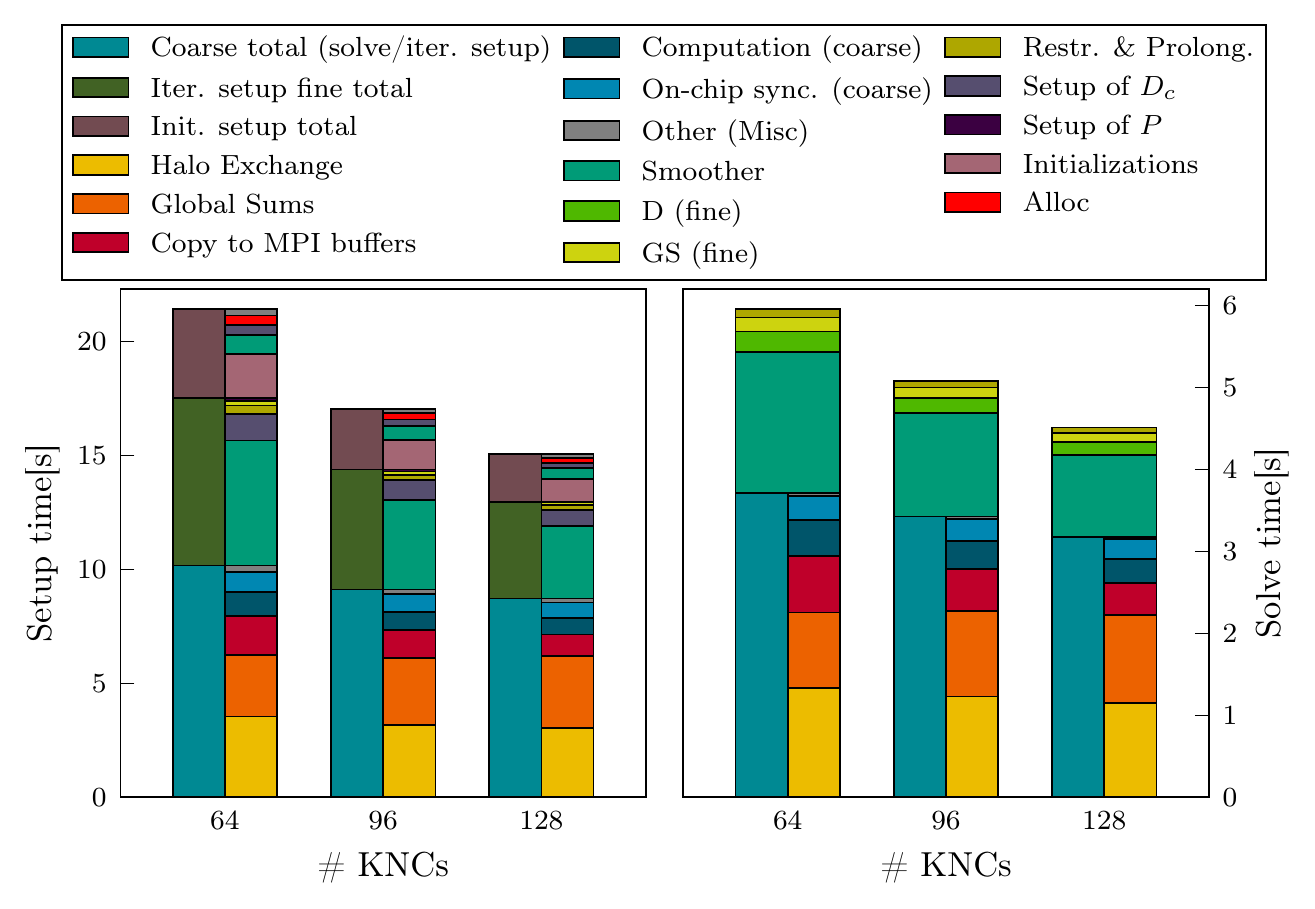}
  \vspace*{-4mm}
  \caption{Strong-scaling plot of the contributions of various MG
    components to the execution time, for a $48^3\times96$ CLS lattice
    with $\beta=3.4$, $m_\pi=220$~MeV, and $a=0.086$~fm.  The
    algorithmic parameters are tuned to minimize the total wall-clock
    time for the 128 KNC run.  The simulations were performed on QPACE 2.}
  \label{fig:setup_solve}
\end{figure}

In Tab.~\ref{tab:perf} we show the speedup obtained by vectorization
of the MG components for a single thread on a single KNC compared to
the original Wuppertal code, for a lattice size of $8^4$ that does not
fit in cache.
\begin{table}
  \centering
  \begin{tabular}{l@{\hspace{3mm}}|@{\hspace{3mm}}c@{\hspace{3mm}}|@{\hspace{3mm}}
    c@{\hspace{3mm}}|@{\hspace{3mm}}c@{\hspace{3mm}}|@{\hspace{3mm}}
    c@{\hspace{3mm}}|@{\hspace{3mm}}c@{\hspace{3mm}}|@{\hspace{3mm}}c}
    MG component & Restrict. & Prolong. & $D_c$ setup & $(D_c)_{i\ne j}$
    & $(D_c)_{ii}$ & GS on aggr.\\ \hline 
    SIMD speedup & 14.1 & 8.6 & 19.7 & 20.2 & 19.5 & 10.8 
  \end{tabular}
  \caption{SIMD speedup for a single thread on a single KNC.}
  \label{tab:perf}
\end{table}
In Fig.~\ref{fig:setup_solve} we show how these improvements affect
the contributions of the various MG components to the total wall-clock
time, for both setup and solve.
We observe that the MG parts that have been optimized (see
Tab.~\ref{tab:perf}) no longer contribute significantly to the
execution time.  The smoother, which was optimized earlier, takes
about 1/3 of the time.  The main contribution now comes from the
coarse-grid solve.  After our optimizations, its computational part
has become cheap so that off-chip communication (halo exchange and
global sums) becomes dominant.
  
We conclude that \WMG\ is a very good target for SIMD architectures.
The run time is now dominated by the communications in the coarse-grid
solve, which will be optimized next.

The code presented here will be made publicly available in the near
future.  We thank Andreas Frommer and Karsten Kahl for helpful
discussions and Daniel Richtmann for producing the figures.

\vspace*{-2mm}

\let\oldbibliography\thebibliography
\renewcommand{\thebibliography}[1]{%
  \oldbibliography{#1}%
  \setlength{\itemsep}{0mm}%
}
\bibliographystyle{JBJHEP_mod}
\bibliography{lat15_amg}

\end{document}